\documentclass[a4paper,fleqn,usenatbib]{mnras}

\usepackage[T1]{fontenc}
\usepackage{ae,aecompl}

\usepackage{graphicx,epstopdf}	
\usepackage{amsmath}	
\usepackage{amssymb}	
\usepackage{multirow}
\usepackage{adjustbox}

\title[Systematic Comparison between Two GD Solutions]{Using Gaia DR2 to make a Systematic Comparison between Two Geometric Distortion Solutions}

\author[Z. J. Zheng et al.]{
Z. J. Zheng,$^{1,2}$
Q. Y. Peng,$^{1}$\thanks{E-mail: tpengqy@jnu.edu.cn} and
F. R. Lin$^{1}$
\\
$^{1}$Sino-French Joint Laboratory for Astrometry, Dynamics and Space Science, Jinan University, Guangzhou 510632, China\\
$^{2}$College of Computer, Guangdong University of Petrochemical Technology, Maoming 525000, China
}

\date{Accepted XXX. Received YYY; in original form ZZZ}

\pubyear{2020}

\begin{document}
\label{firstpage}
\pagerange{\pageref{firstpage}--\pageref{lastpage}}
\maketitle

\begin{abstract}
Gaia Data Release 2~(Gaia DR2) provides high accuracy and precision astrometric parameters~(position, parallax, and proper motion) for more than 1 billion sources and is revolutionising astrometry. For a fast-moving target such as an asteroid, with many stars in the field of view that are brighter than the faint limit magnitude of Gaia~(21 Gmag), its measurement accuracy and precision can be greatly improved by taking advantage of Gaia reference stars. However, if we want to study the relative motions of cluster members, we could cross-match them in different epochs based on pixel positions. For both types of targets, the determination of optical field-angle distortion or called geometric distortion~(GD) in this paper is important for image calibration especially when there are few reference stars to build a high-order plate model. For the former, the GD solution can be derived based on the astrometric catalog's position, while for the latter, a reference system called `master frame' is constructed from these observations in pixel coordinates, and then the GD solution is derived. But are the two GD solutions in agreement with each other?

In this paper, two types of GD solutions, which are derived either from the Gaia DR2 catalog or from the self-constructed master frame, are applied respectively for the observations taken by 1-m telescope at Yunnan Observatory. It is found that two GD solutions enable the precision to achieve a comparable level~($\sim$10 mas) but their GD patterns are different. Synthetic distorted positions are generated for further investigation into the discrepancy between the two GD solutions. We aim to find the correlation and distinction between the two types of GD solutions, and their applicability in high precision astrometry.
\end{abstract}

\begin{keywords}
astrometry -- reference system -- methods: data analysis -- techniques: image processing
\end{keywords}



\section{Introduction} \label{sec:intro}
Astrometry aims to measure precisely and accurately the positions and motions of celestial objects. For photographic or CCD astrometry, a typical practice is to determine the relationship between objects' equatorial coordinates and their pixel/measurement coordinates in the image in the focal plane. However, if we are only interested in small changes in the positions of celestial objects, such as relative parallaxes or proper motions of the objects, ``we need only determine changes in the position of the object  over a period of months or years and the position of the object is of secondary importance''~\citep{Altena2013}. It is adequate to solve the relative parallaxes or proper motions using the pixel/measurement coordinates directly. We refer to the practice as differential astrometry while the former as photographic astrometry in this paper.

In the past, photographic astrometry had limitations, such as only few reference stars being available or some systematic errors for them. A well-known example is the astrometry of Hubble Space Telescope~(HST), for its power of viewing much fainter magnitude~($\gg$21 mag), beyond the Gaia detection limit. In most cases, differential astrometry might be a better option for HST, which can maintain the precision of the pixel coordinates~(0.01 pixel, $\sim$0.5 mas for PC chip and $\sim$1 mas for WF chip). Differential astrometry is also favored by ground-based telescopes for high precision, such as the determination of relative proper motion~\citep{Bellini2010_2}, parallax~\citep{Sahlmann2016} or the orbit of binary systems~\citep{Ginski2013}.

Now, the ESA space mission, Gaia, will revolutionize astrometry in the coming decades. It provides absolute positional measurements over the entire sky with unprecedented accuracy and magnitude completeness. Many ground-based telescopes must benefit from the catalog to achieve far higher precision astrometry. For deep, ground- or space-based telescopes, the potential of using Gaia DR2 as a calibration tool is also tremendous, especially for many projects such as satellite or asteroid orbit determination. Uncertainties in the positioning for a high signal-to-noise~(SNR) star image is usually 1$\%$ pixel~(2 mas, see Table.1 in~\citealt{Libralato2014}), and it is not superior to the typical precision in Gaia's faint end, 0.7 mas at G=20 mag. So in addition to differential astrometry using only pixel coordinates of stars, large telescopes can also use Gaia reference stars with high SNR images to directly link the pixel positions with the celestial positions. For the forthcoming wide-fast-deep survey by Large Synoptic Survey Telescope~(LSST), which is dedicated to carrying out high-quality astrometry, Gaia would also serve as a synergy to improve its measurement for proper motion of sources fainter than 21 Gmag~\citep{Casettidinescu2018Astrometry}.

For both types of astrometry mentioned above~(photographic astrometry and differential astrometry), pixel positional measurement~(Point Spread Function centering or two-dimensional Gaussian centering) and calibrating for optical field-angle distortion are important for reaching the intrinsic error floor of the observations. In this paper, the optical field-angle distortion, also called geometric distortion~(GD) is focused on. Although GD can be expressed by a high-order polynomial if there are sufficient Gaia stars in the field of view~(FOV), a GD pattern of the CCD camera in a certain time is still desirable. For example, when performing Synthetic Tracking~\citep{Zhai2018AJ} for detecting and tracking near-Earth objects~(NEO), it is convenient and timesaving to transform the pixel positions to the distortion-free positions by utilizing GD solution, instead of building a high order distortion model~(Equations 3 and 4 in~\citealt{Zhai2018AJ}).

For both types of astrometry, there are corresponding GD solutions. For differential astrometry, a distortion-free reference frame called the master frame is constructed by the observations themselves in pixel coordinate and taken as a reference to derive the GD solution~\citep{Anderson2006}. The solution has been successfully applied for the astrometry of HST and several ground-based telescopes~\citep{Anderson2006,Bellini2010,Libralato2014}. For photographic astrometry, we proposed an alternative GD solution, taking an astrometric catalog as reference~\citep{Peng2012} and the solution is also successfully applied for the astrometry of some natural satellites~\citep{Peng2015MNRAS,Peng2017,Wang2017}. However, if derived from the same observations, are the two GD solutions in agreement with each other? Further, are the positions after GD correction in agreement with the ones of an astrometric flat field?

In this paper, we compare the GD solution based on photographic astrometry~(hereafter GD1), which takes positions from Gaia DR2 as a reference, with the GD solution based on differential astrometry~(hereafter GD2), which takes positions from the master frame as a reference. And further analysis is made for the two GD patterns through simulation.

The paper is organized as follows: in Sect.2, we briefly describe the instrument and the observations. In Sect.3, we derive the two GD solutions and compare their patterns. In Sect.4, we compare the measurement in terms of precision and accuracy. In Sect.5, through simulation, we further investigate the non-linear terms of the two GD solutions. We also discuss if the positions of the master frame are in agreement with the International Celestial Reference System~(ICRS). The final section summarizes our results.

\section{Instrument and observations} \label{sec:ins&obs}

The open cluster M35 was observed near the zenith to minimize atmosphere refraction effects with the 1-m telescope at Yunnan Observatory~(YNAO). To derive the GD solution, a dithering strategy was arranged by a set of pointings with a step of $\sim$1$'$ and covered a sky region of $\sim$23$'$$\times$ 23$'$. The observations were obtained in I filter with Andor's iKon-XL 231 CCD camera. Details of the telescope and CCD camera are shown in Table~\ref{Tab TelsSpec}, and observations are shown in Table~\ref{Tab ObsSpec}. For simplicity, the three observation-sets are called Obs1~(2018.11.11), Obs2~(2018.11.12) and Obs3~(2018.11.13) respectively.

\begin{table}
\caption{Specifications of the telescope and CCD camera.}
\centering
\begin{tabular}{@{}cc@{}}
\hline\hline
Focal length & 13.3 m \\
 Diameter of primary mirror  & 1 m \\
 FOV  & 16$'$$\times$16$'$ \\
 Size of pixel   & $15{\times}15$ $\mu$m${^2}$\\
 Size of CCD array  & $4096{\times}4112$ \\
 Approximate pixel scale  & $0\farcs234$/pix  \\
\hline
\end{tabular}
\label{Tab TelsSpec}
\end{table}

\begin{table}
\caption{Specifications of the observations.}
\centering
\begin{tabular}{@{}cccccc@{}}
\hline\hline
Index &Date  & Exp-time  & Mean Seeing & Airmass  
\\
& & & (arcsec) & (sec~$z$)
\\\hline

1 &2018.11.11    & $55{\times}60s$   &
1.90 & $1.01-1.10$ \\
2 &2018.11.12    & $43{\times}60s$   &
1.85 & $1.00-1.05$ \\
3 &2018.11.13    & $52{\times}60s$   &
 1.68 & $1.00-1.04$ \\
\hline
\end{tabular}
\label{Tab ObsSpec}
\end{table}

\section{Deriving the GD solution}\label{sec:gdsolution}

\begin{figure}
\includegraphics[width=\columnwidth]{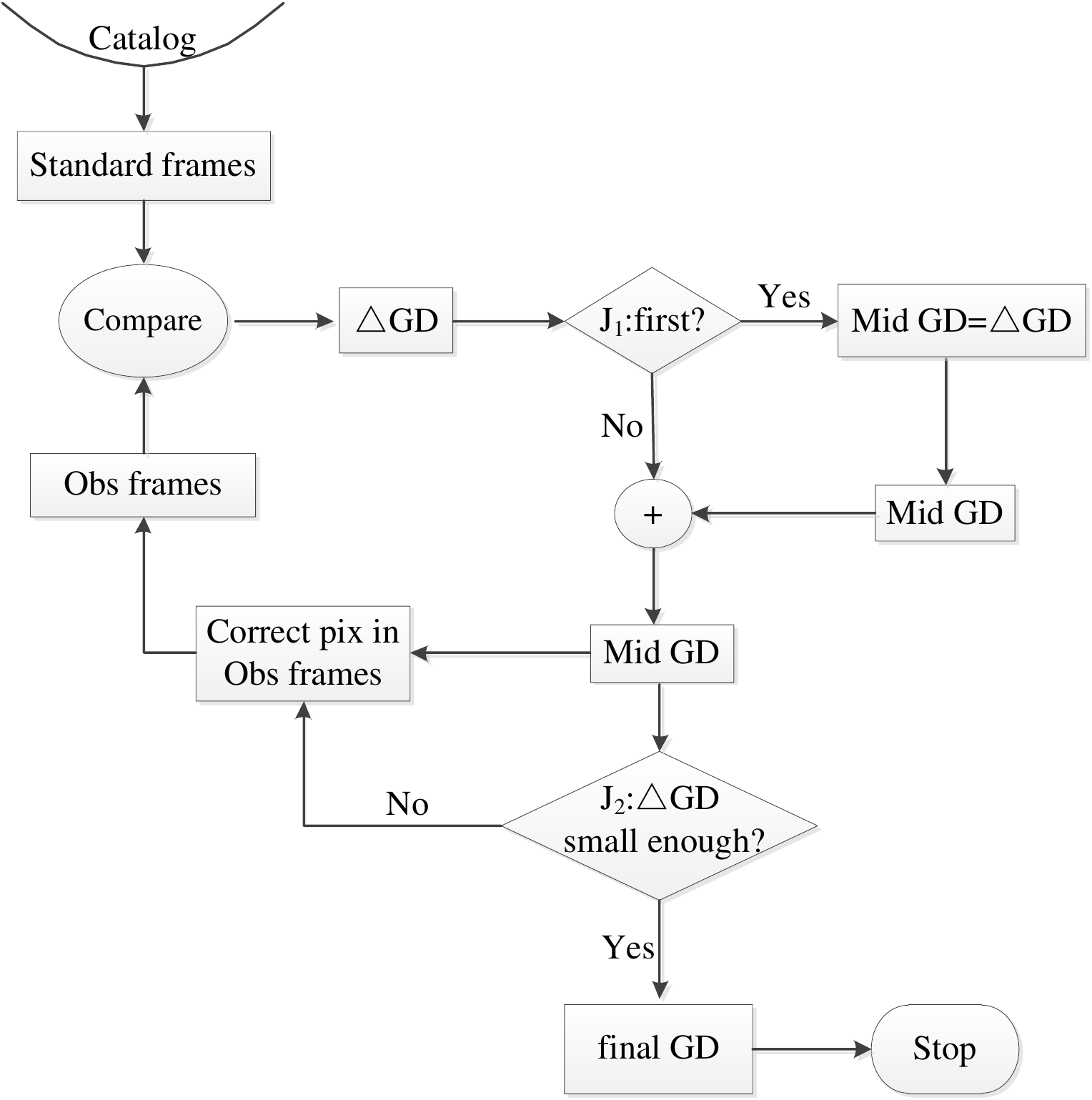}
\caption{Diagram of GD1 procedure.}
\label{Fig9}
\end{figure}

\begin{figure}
\includegraphics[width=\columnwidth]{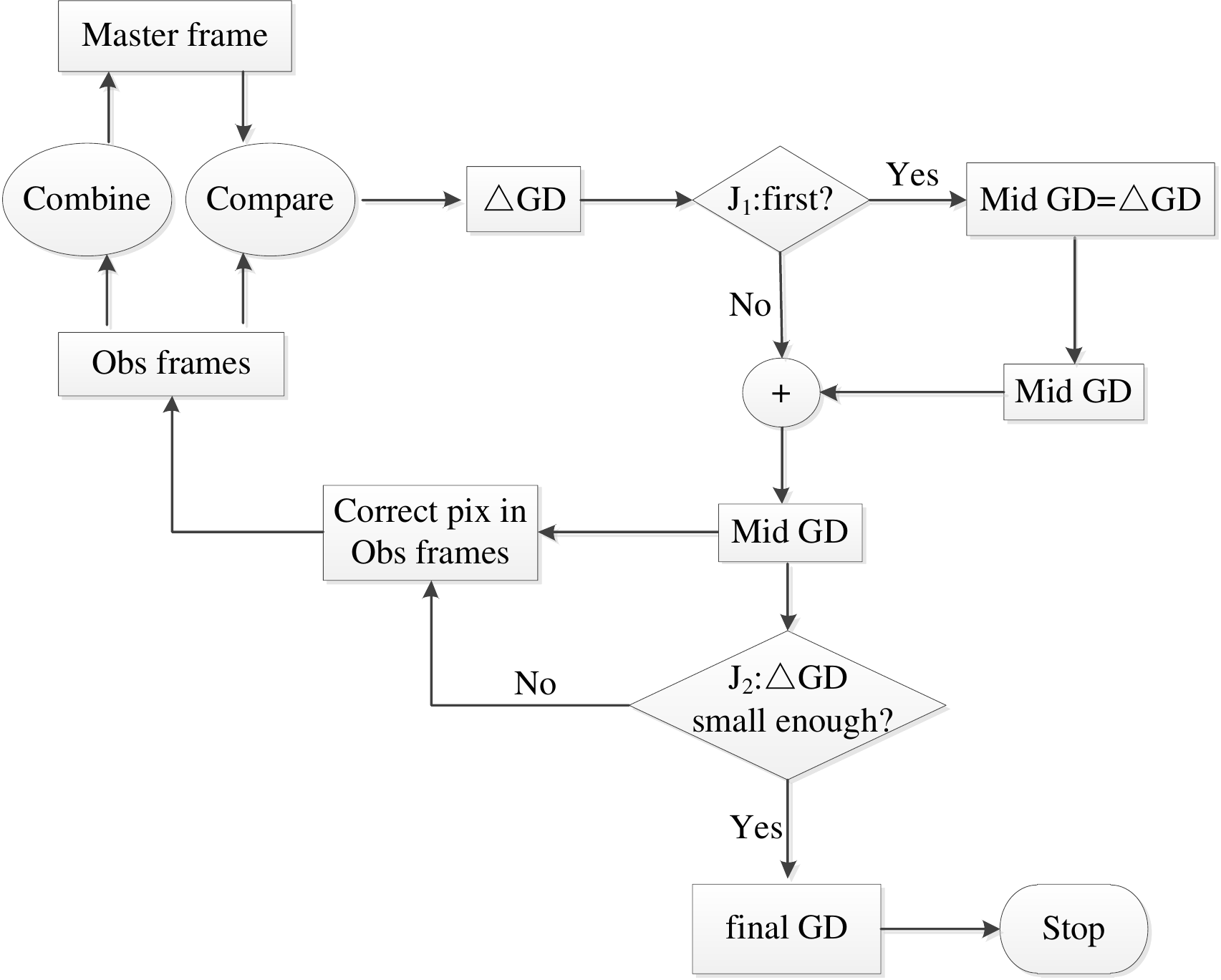}
\caption{Diagram of GD2 procedure.}
\label{Fig10}
\end{figure}

The most straightforward way to solve for GD should be to compare the observations with a distortion-free reference frame (DFF). The residuals would be the direct depiction of the GD effect in the observation. DFF is defined as a coordinate system that any observed frame after GD correction can be transformed to, by just a 4-parameter conformal transformation (i.e. considering translation, rotation and scaling). However, the actual DFF does not exist but we could build an approximate one, based on the astrometric catalog or the observations themselves. For photographic astrometry, we can take the tangential standard coordinate called the standard frame for each exposure as a reference, which is calculated from an astrometric catalog, considering general astrometric effects~(parallax, aberration, atmosphere refraction, etc). And for differential astrometry, the reference frame called the master frame is combined by a set of average positions from the observations~\citep{Anderson2003}. The procedures of the two GD solutions are visualized on Fig.~\ref{Fig9} and Fig.~\ref{Fig10} respectively. The procedure is interpreted in the following section.

\begin{description}

\item[$\bullet$] Catalog and Standard frame: GD1 is modeled based on the standard frame, i.e. the tangential standard coordinates calculated by gnomonic projection from the astrometric catalog --- Gaia DR2. For each observed frame, we calculate its corresponding standard frame, considering general astrometric effects.

\item[$\bullet$] Master frame: GD2 is modeled based on the master frame, of which the positions are constructed from observed frames, to maintain the accuracy and precision of the pixel coordinates. The master frames are constructed for each observation-set. In the following, the construction of the master frame is given in detail.

To build the master frame, we find a conformal transformation from each observed frame to the centermost frame by the least square method, determined from their common stars' pixel positions. The position of each star on the master frame is calculated from the $\sigma$-clipped average of all the transform positions from the same stars of different frames. And the RMS~(marked as $\sigma$) of each position of the master frame is served as a weight for transformation in the next iteration. At the same time, a star on the frame would be flagged if it has a transform outlier during the estimation for the master frame's position, so that it would not be used in the next iteration. Also, new stars are added to the master frame, and finally, the master frame covers all the observed stars. The primitive master frame is then used to improve the transformation, with the weights and flags determined by the previous iteration. The procedure iterates by several times until the positions on the master frame converge to 0.001 pixel~($\sim$0.2 mas) in each direction.

\item[$\bullet$] Compare: The observed frame's positions would be compared with the DFF's, using a 4-parameter conformal transformation. The transformation residuals of each star would reveal the GD effects to some extent, though mixing up with other errors~(catalog error for GD1, transformation model error and centering error for both). Thanks to the dithering strategy, the same star is imaged in various positions of the CCD, and the transformation residual discloses the GD effect as a function of pixel position. Specifically, the catalog error can be canceled out when relating with the error of the same star observed in different frames~(refer to~\citeauthor{Peng2012} 2012 in detail).

\item[$\bullet$] GD modeling and Correction: both GD solutions are presented simply by a look-up table which is constituted by 16${\times}$16 bins~(each bin is made up of 256${\times}$257 pixels) over the CCD chip. This setting proves to be the best compromise between the need for an adequate number of bins to model the GD solution and adequate sampling of stars in each bin.

The GD effect in the center of each bin is estimated from the transformation residuals of all the stars in the bin, through a $\sigma$-clipped average solution. After the first iteration, a temporary file, \textit{Mid GD} is used to store the current GD solution and the positions of stars on each frame are corrected by the bi-linear interpolation of GD estimations of four closest bins. In the next iteration of the GD solution, the remanent GD effects of each bin are solved again and they are added to \textit{Mid GD} file. The GD solution iterates until the adjustment in each bin is less than 0.001 pixel. It should be noted that, for GD2 solution, the master frame would be reconstructed based on the corrected positions in each iteration.
\end{description}

The above reduction would be performed after the standard procedures including de-bias and flat fielding. And the pixel positions of stars are achieved by a two-dimensional Gaussian centering.

\begin{figure*}
    \begin{minipage}{\textwidth}
    \centering
	\includegraphics[width=0.41\columnwidth]{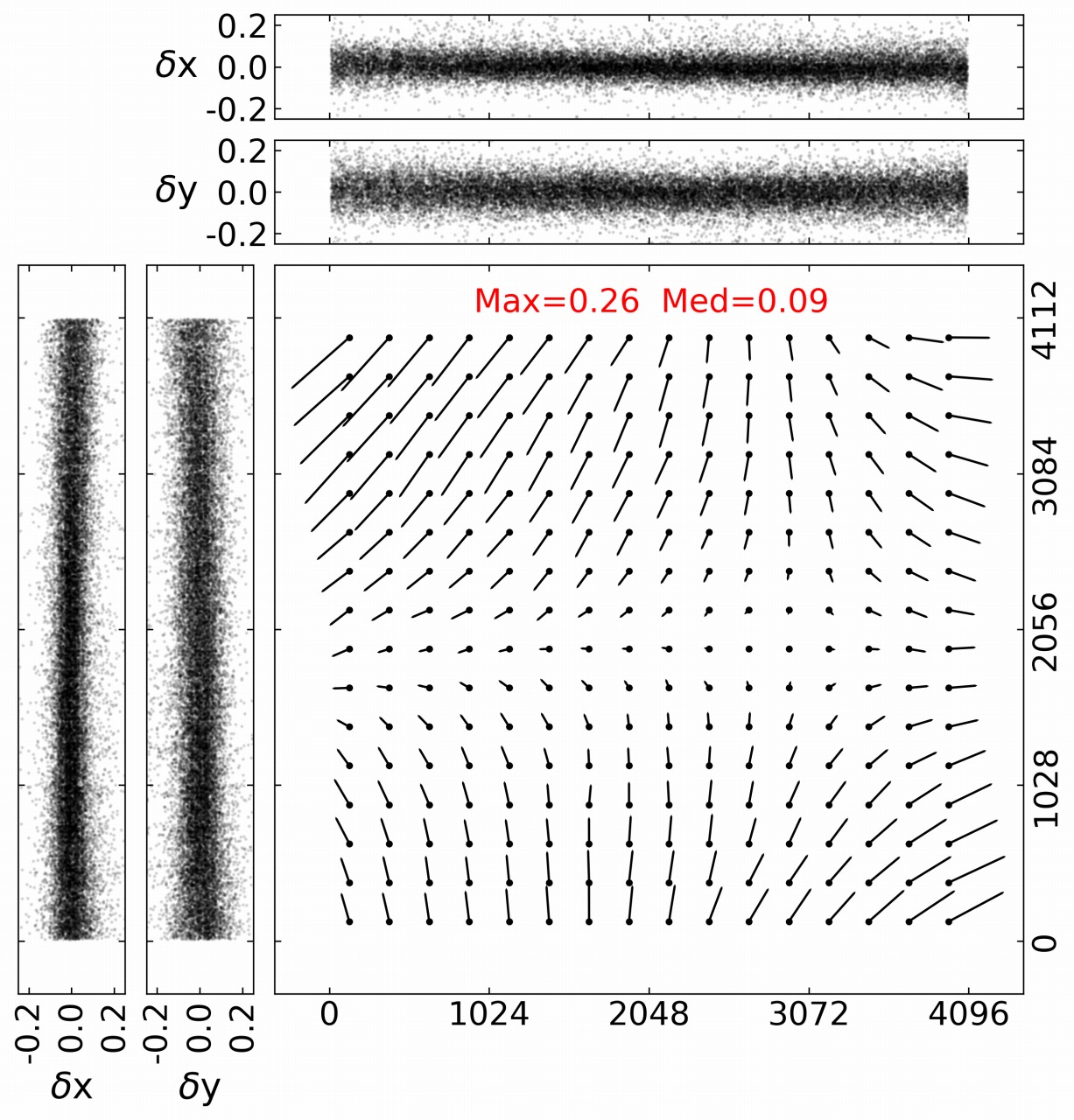}
	\includegraphics[width=0.41\columnwidth]{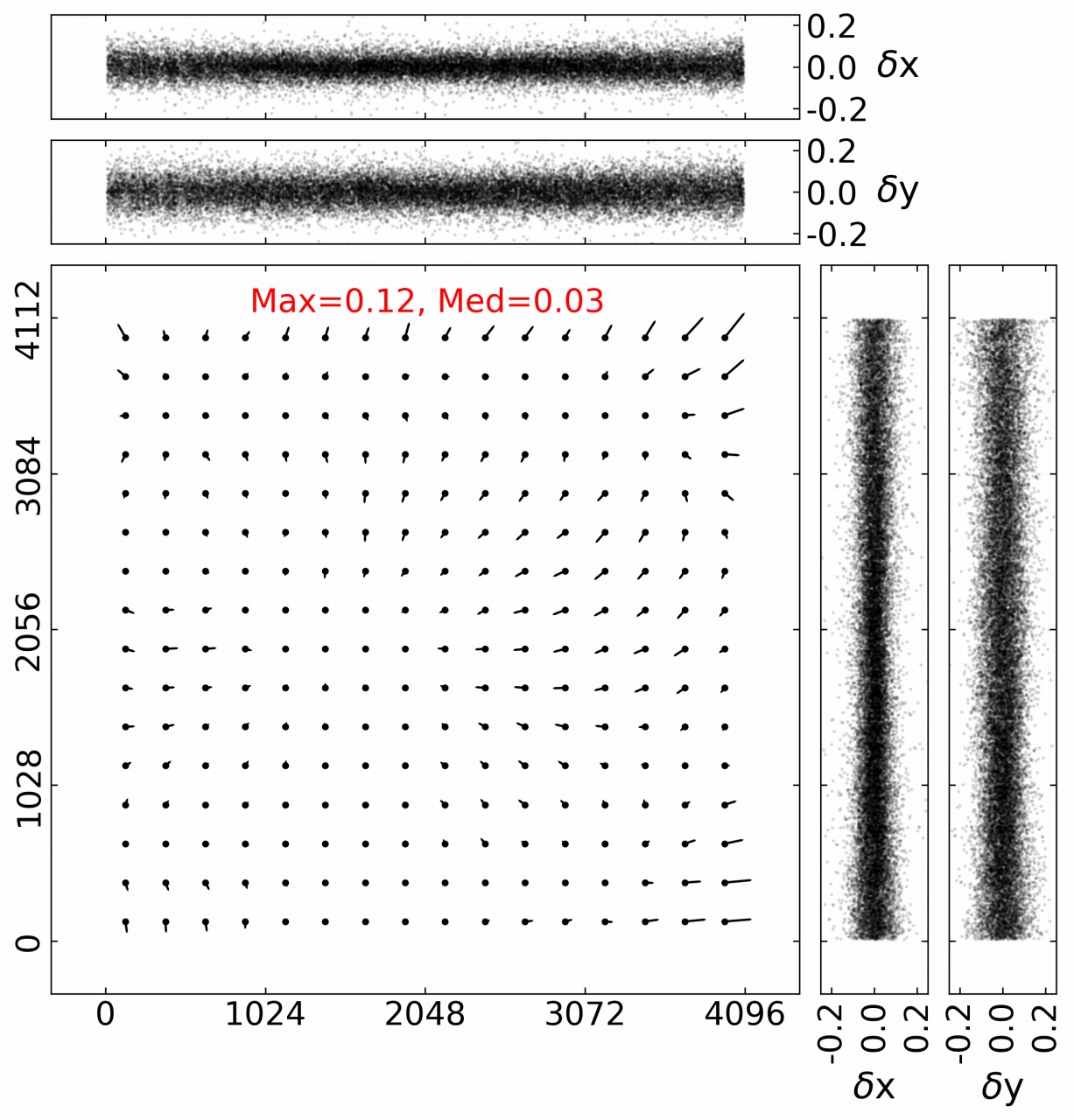}
    \end{minipage}
    \vspace{0in}
    \begin{minipage}{\textwidth}
    \centering
	\includegraphics[width=0.41\columnwidth]{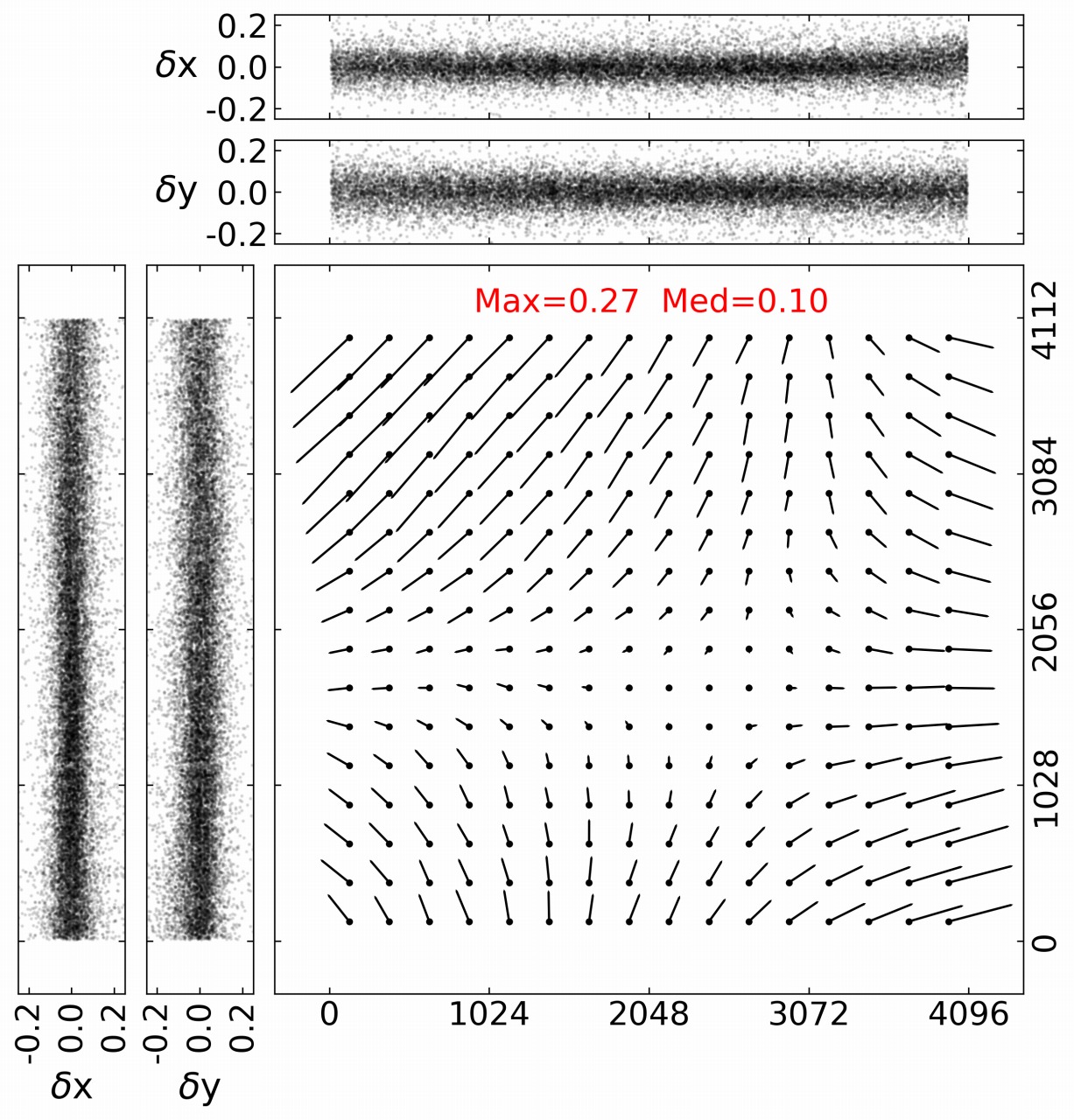}
	\includegraphics[width=0.41\columnwidth]{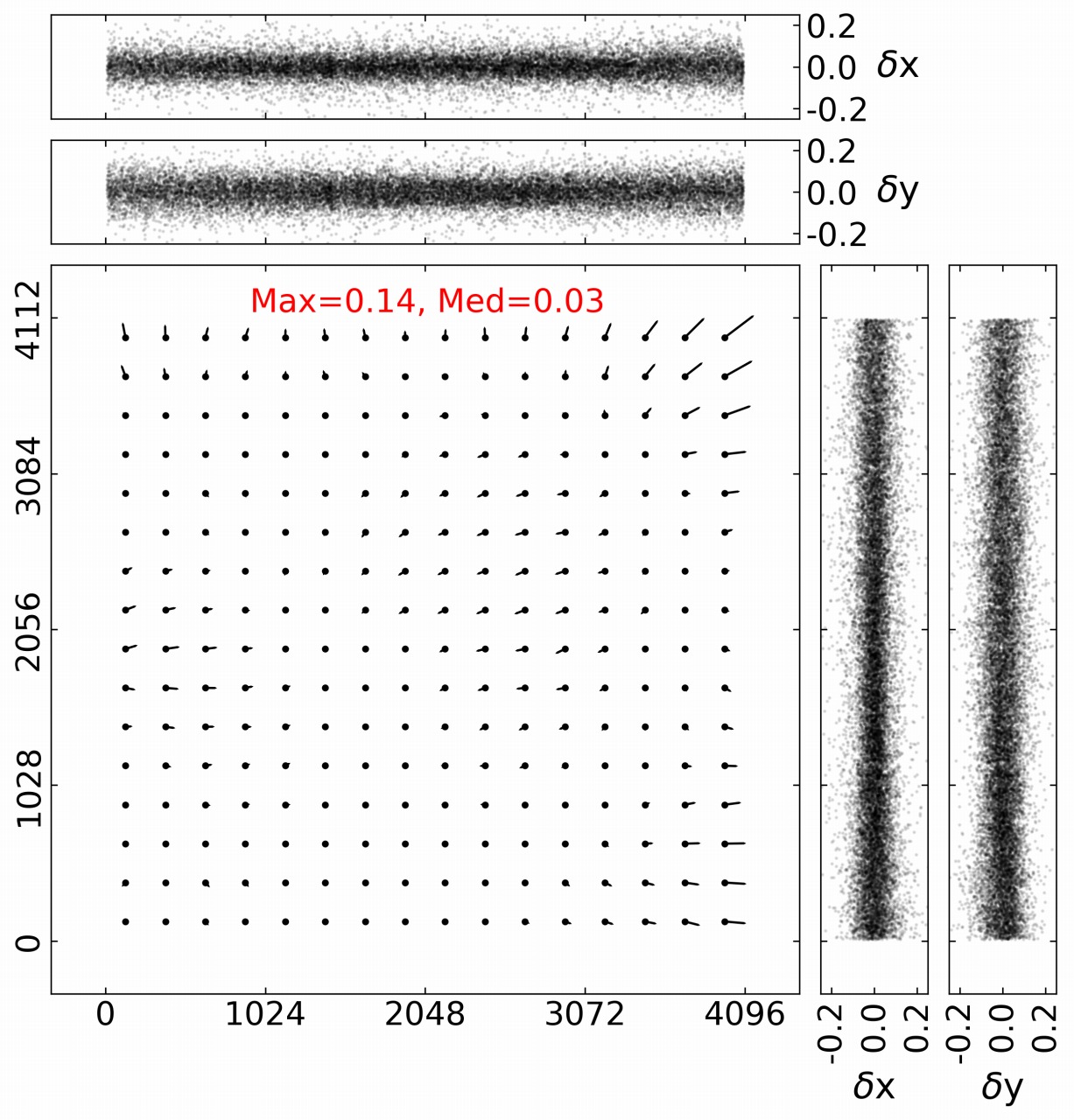}
    \end{minipage}
    \vspace{0in}
    \begin{minipage}{\textwidth}
    \centering
	\includegraphics[width=0.41\columnwidth]{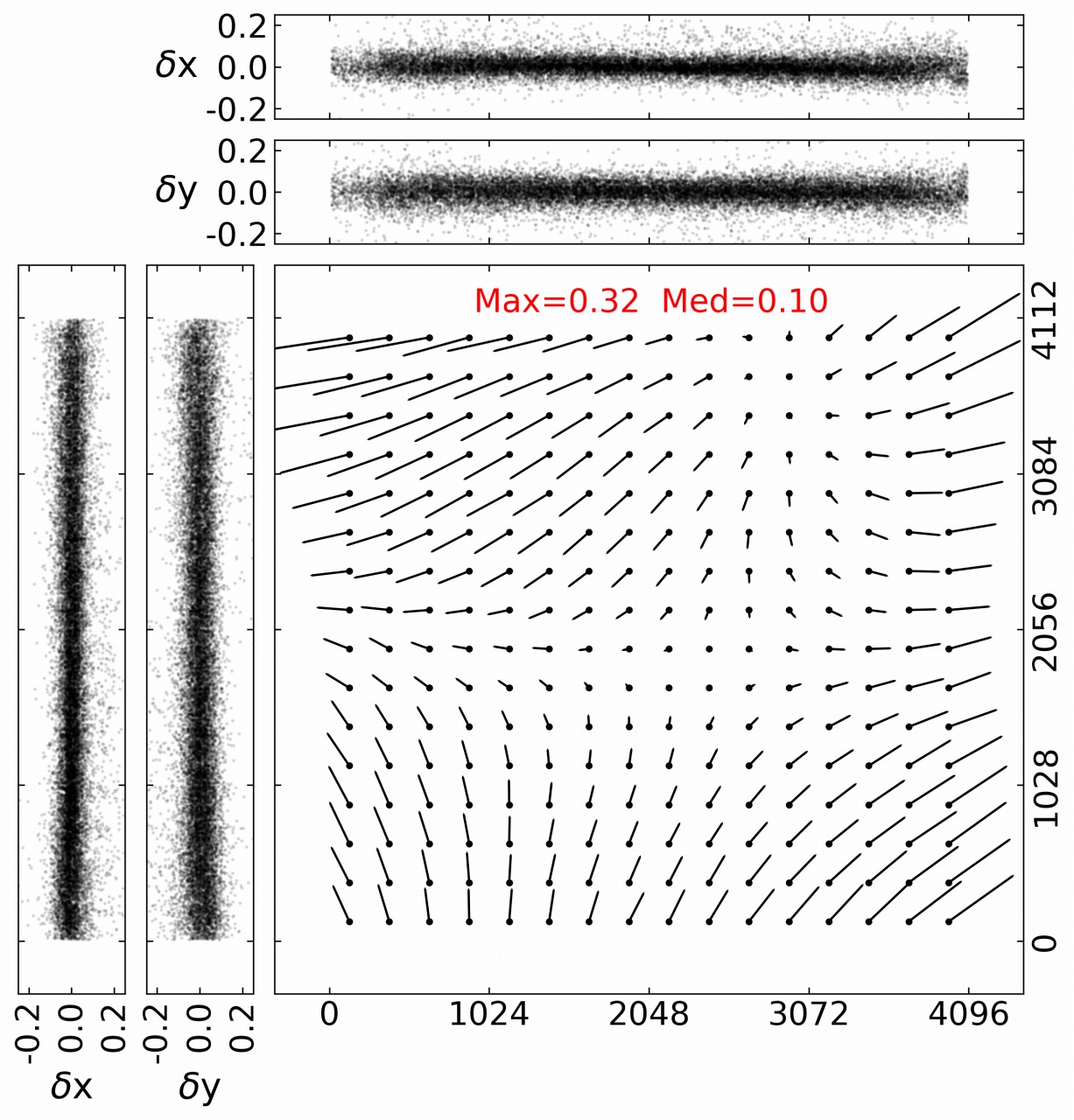}
	\includegraphics[width=0.41\columnwidth]{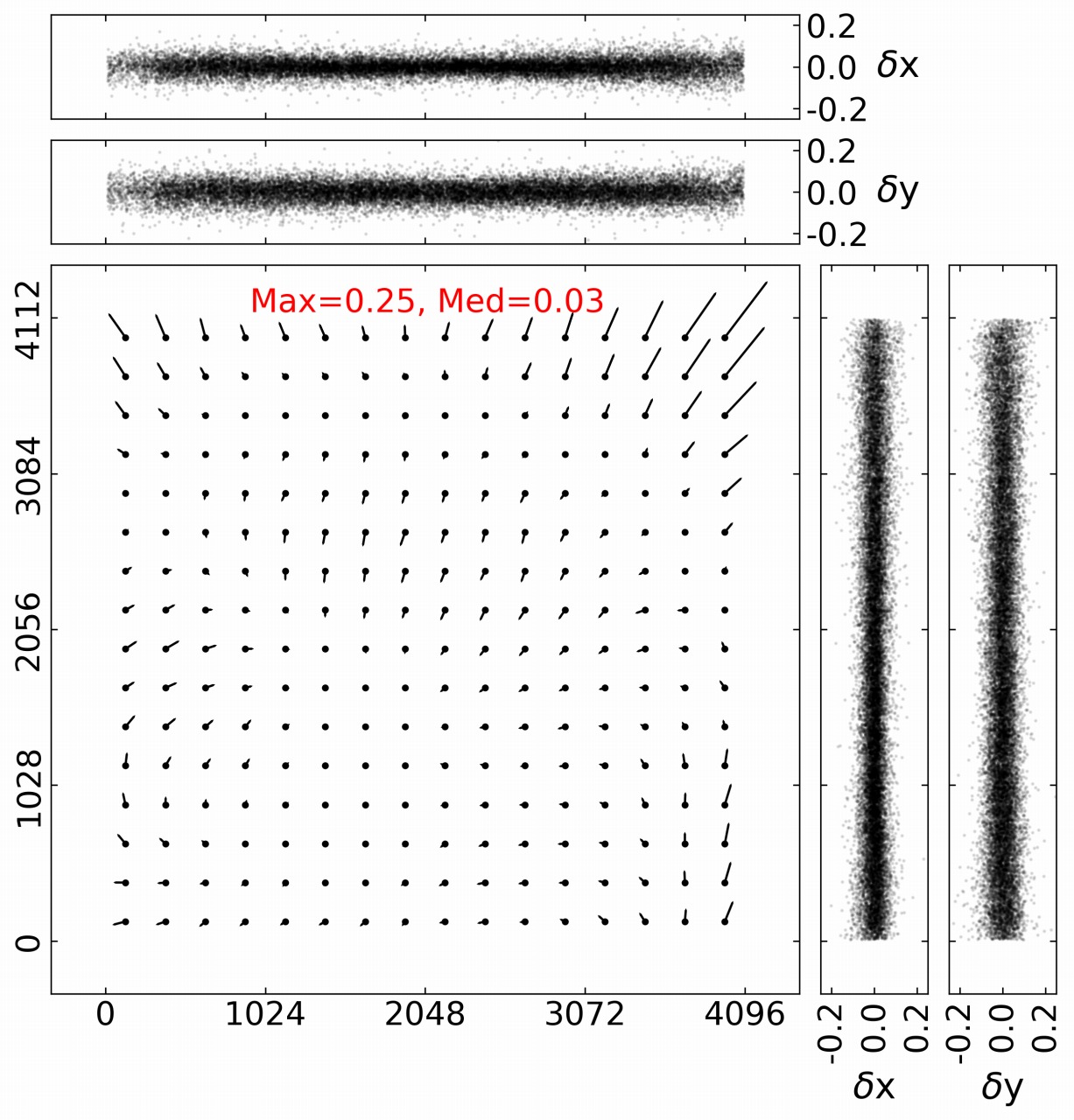}
    \end{minipage}
    \caption{GD1~(left) and GD2~(right) patterns for Obs1~(top), Obs2~(middle) and Obs3~(bottom) and residual trends along X and Y axes for stars brighter than 16 Gmag. Unit is expressed in pixel. Maximums and medians are shown at top of each panel. And the vectors are magnified by a factor of 2000.}
    \label{Fig1}
\end{figure*}

We derive two types of GD solutions for each observation-set and perform the correction. The distortion patterns have been smoothed with a 5 $\times$ 5 Gaussian smoothing kernel, and we have verified that the smoothing did not compromise our solution. For simplicity, we show the patterns of the two GD solutions and the transformation residuals of bright stars~(Gmag$<$16) after GD correction along with $X$ and $Y$ axes in Fig.~\ref{Fig1}. For each observation-set, two GD patterns are different from each other. GD1 exhibits a considerable linear skew while GD2 little. And the magnitude of GD1 is much larger than GD2. The GD pattern of Obs3 is very different from Obs1 and Obs2. It may result from the maintenance during the day since the CCD is just put into operation for only one month. We test the stability of the solution in the next section.

\section{Comparison in terms of measurement precision and systematic errors}

In this section, we compare the two GD solutions in terms of the measurement precision and systematic errors by inter-comparison between datasets.

\subsection{Comparison of precision}

After GD correction, a conformal transformation is used to bring each star's position into the standard frame or the master frame. We consider $\sigma$~($\sqrt{\sigma_{\rm{\alpha cos\delta}}^{\rm{2}}+\sigma_{\rm{\delta}}^{\rm{2}}}$ for photographic astrometry and $\sqrt{\sigma_{x}^{2}+\sigma_{y}^{2}}$ for differential astrometry) of the transformation residuals of the same star, to evaluate the accuracy of the GD solutions. To facilitate the comparison, $\sigma$ for GD2 is converted to arcsecond unit with an average pixel scale $0\farcs234$/pix. For all the observations, the pixel scale is very stable at the mas level.

\begin{figure*}
\includegraphics[width=\textwidth]{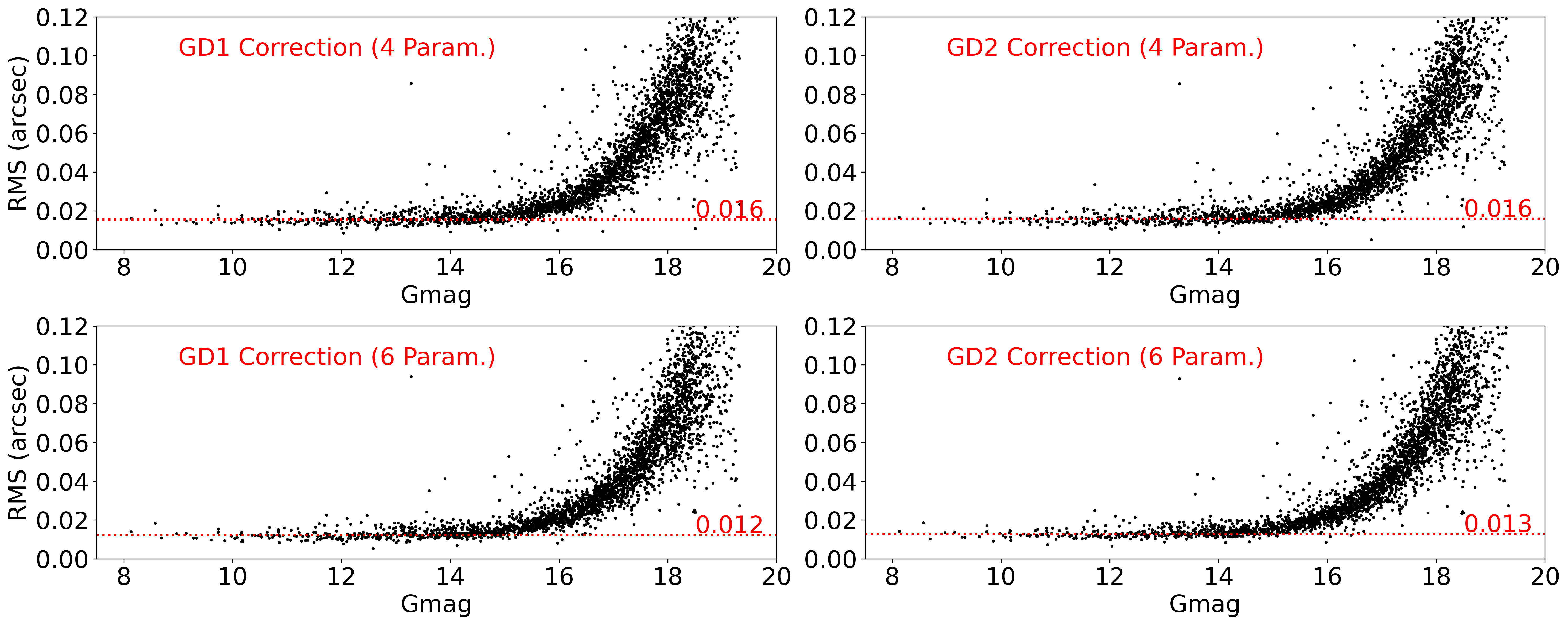}
\caption{Top left: After GD1 correction for Obs1, RMS~($\sqrt{\sigma_{\rm{\alpha cos\delta}}^{\rm{2}}+\sigma_{\rm{\delta}}^{\rm{2}}}$) as a function of Gmag by a four-parameter conformal transformation to the standard frames. Top right: After GD2 correction for Obs1, RMS~($\sqrt{\sigma_{x}^{2}+\sigma_{y}^{2}}$) as a function of Gmag by a four-parameter conformal transformation to the master frame. Bottom left: After GD1 correction, RMS of using a six-parameter linear transformation. Bottom right: After GD2 correction, RMS of using a six-parameter linear transformation. The red dashed line marks the median values of the stars brighter than 14 Gmag.}
\label{Fig2}
\end{figure*}

\begin{figure*}
\includegraphics[width=\textwidth]{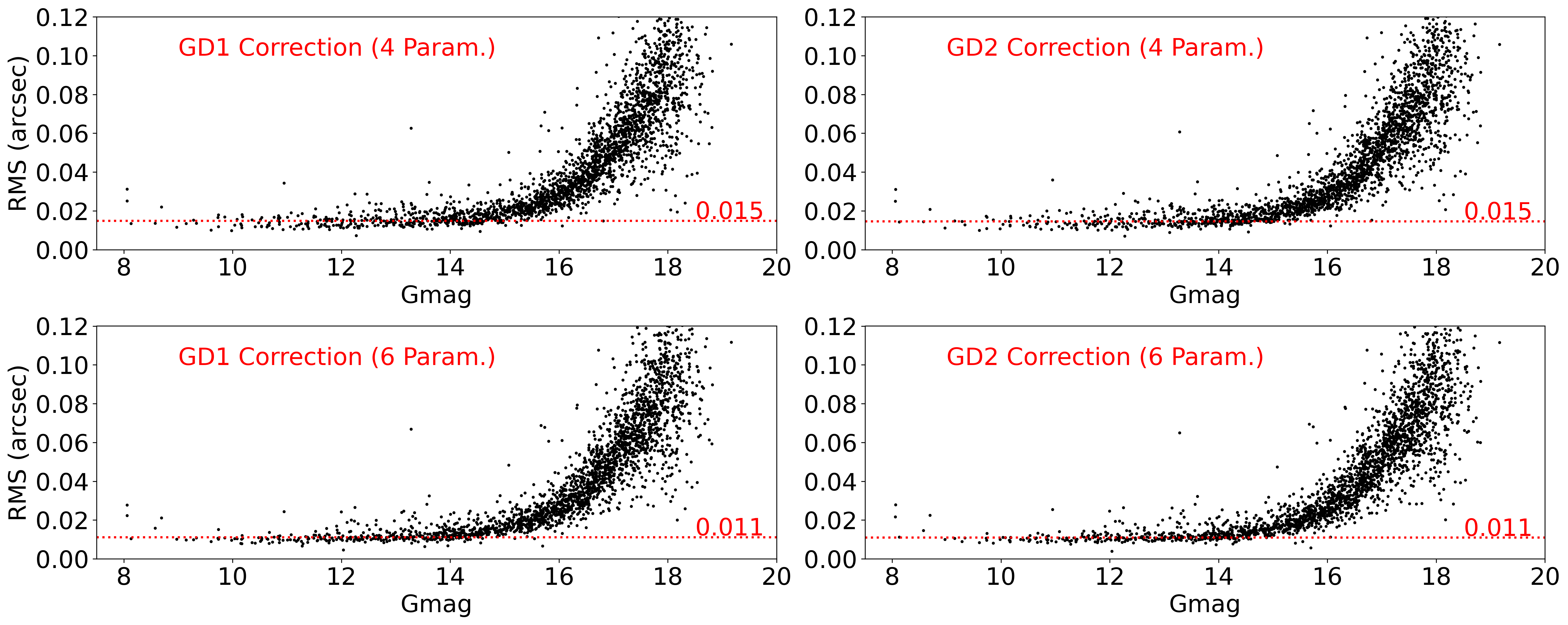}
\caption{Similar to Fig.~\ref{Fig2} but for Obs2.}
\label{Fig3}
\end{figure*}

\begin{figure*}
\includegraphics[width=\textwidth]{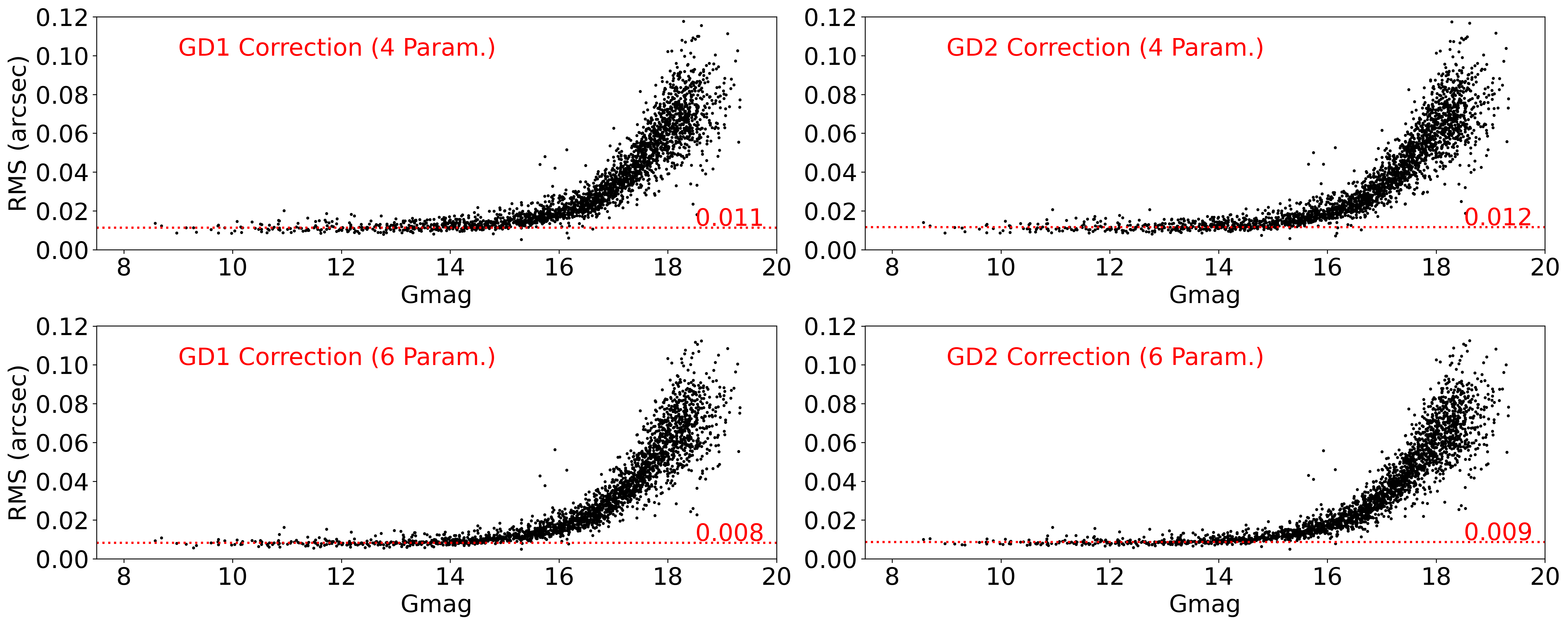}
\caption{Similar to Fig.~\ref{Fig2} but for Obs3.}
\label{Fig4}
\end{figure*}

The final results are shown as function of Gaia's Gmag in the top two subfigures of Fig.~\ref{Fig2}, Fig.~\ref{Fig3} and Fig.~\ref{Fig4} for three observation-sets respectively. A dashed red line indicates the median precisions of stars brighter than 14 Gmag. It is found that two GD
solutions enable the precision to achieve a comparable level.

It should be noted that the derived solution is an average GD solution for the telescope during the observation run. The linear terms of distortion are easily changed by insufficient correction of atmosphere refraction for GD1, or differential atmosphere refraction for GD2~(though near the zenith), and variation of telescope's flexure. Taking into account these errors, a 6-parameter linear transformation is applied instead and the results are shown in the bottom two subfigures of Fig.~\ref{Fig2}, Fig.~\ref{Fig3} and Fig.~\ref{Fig4}. Similar GD correction capacities are found for the two solutions.

To test the GD stability, we also look into the precisions after applying cross-correction to the observations. That is, for each observation-set, we also adopt the GD solution which is derived from other observation-set. Only the median values of the precision for the stars brighter than 14 Gmag after correction is listed in Table~\ref{Tab Stable}. The largest precision loss is 2 mas for Obs2 when GD solution is derived from Obs3 and the transformation model is a linear transformation. The other precision losses are within 1 mas. For high-precision astrometry, we suggest deriving the GD solution from the observation-set itself, as it is continuously evolving.

\begin{table}
\caption{The median values of the precision in arcsecond for the stars brighter than 14 Gmag after applying GD solutions derived from different observation-sets. The first column indicates the corrected observation-set and the referred GD solution. The last four columns are the precisions after applying GD1 and GD2 solutions respectively, through different transformation models.}
\centering
\begin{tabular}{ccccc}
\hline
\multirow{2}*{Obs/Ref} & \multicolumn{2}{c}{Conformal} & \multicolumn{2}{c}{Linear} \\
&GD1  & GD2&GD1  & GD2\\\hline
Obs1/Obs1 & 0.016  & 0.016 & 0.012  & 0.013 \\
Obs1/Obs2 & 0.016  & 0.016 & 0.013  & 0.013 \\
Obs1/Obs3 & 0.016  & 0.016 & 0.013  & 0.013 \\\hline
Obs2/Obs2 & 0.015  & 0.015 & 0.011  & 0.011 \\
Obs2/Obs1 & 0.015  & 0.015 & 0.011  & 0.012 \\
Obs2/Obs3 & 0.015  & 0.015 & 0.013  & 0.013 \\\hline
Obs3/Obs3 & 0.011  & 0.012 & 0.008  & 0.009 \\
Obs3/Obs1 & 0.012  & 0.012 & 0.009  & 0.009 \\
Obs3/Obs2 & 0.012  & 0.012 & 0.009  & 0.009 \\
\hline
\end{tabular}
\label{Tab Stable}
\end{table}

\subsection{External and internal check for the accuracy}
As a relative self-calibration technique, GD2 solution provides the master frame, which can be presented as a provisional reference frame, not only demanding high-precision but high-accuracy as well. The internal precision is a metric for relative measurement, however, it is an internal estimate for the errors. In case of missing out hidden systematic errors, we would compare the master frames derived from different observation-sets for an external check. For comparison, an internal check is also done for the GD1 solution. It should be noted that we only considered stars brighter than 16 Gmag, which are observed at least 5 times.

\begin{figure*}
\includegraphics[width=\textwidth]{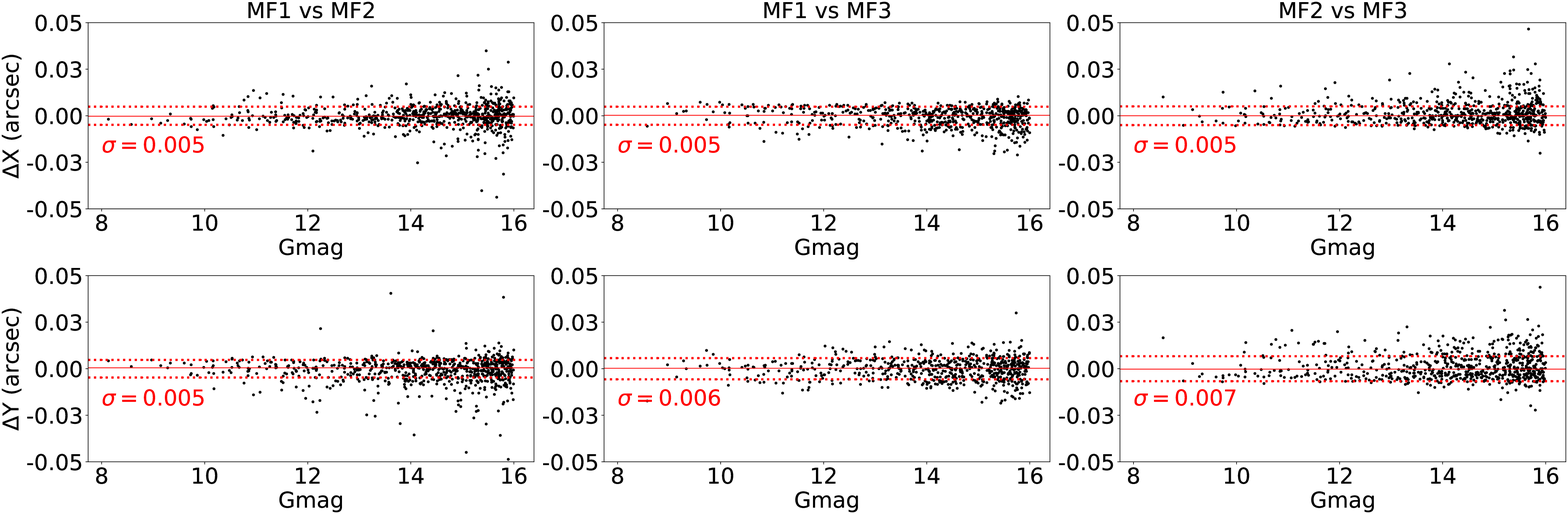} \\
\caption{The residuals along $X$ or $Y$ axes when two master frames~(MF for short) relate with each others. Left:~MF of Obs1 relates with MF of Obs2. Middle:~MF of Obs1 relates with MF of Obs3. Right:~MF of Obs2 relates with MF of Obs3. The red dashed line is marked at $\pm\sigma$ of the residuals and the solid line is marked at average value.}
\label{Fig7}
\end{figure*}

Since the variation of thermal-induced or flexure-induced in telescope's optics and differential refraction, the linear terms would easily change during the observation. As a result, we just explore non-linear terms between the master frames. Ignoring the proper motion effects of stars in one or two days, a linear transformation is applied to relate two master frames. The transform residuals along $X$ or $Y$ axes are shown in Fig.~\ref{Fig7}. Note that the results have been converted to arcsecond unit with the approximate scale $0\farcs234$/pix. The $\sigma$ is stable within about 6 mas at average in each direction.

\begin{figure*}
\includegraphics[width=\textwidth]{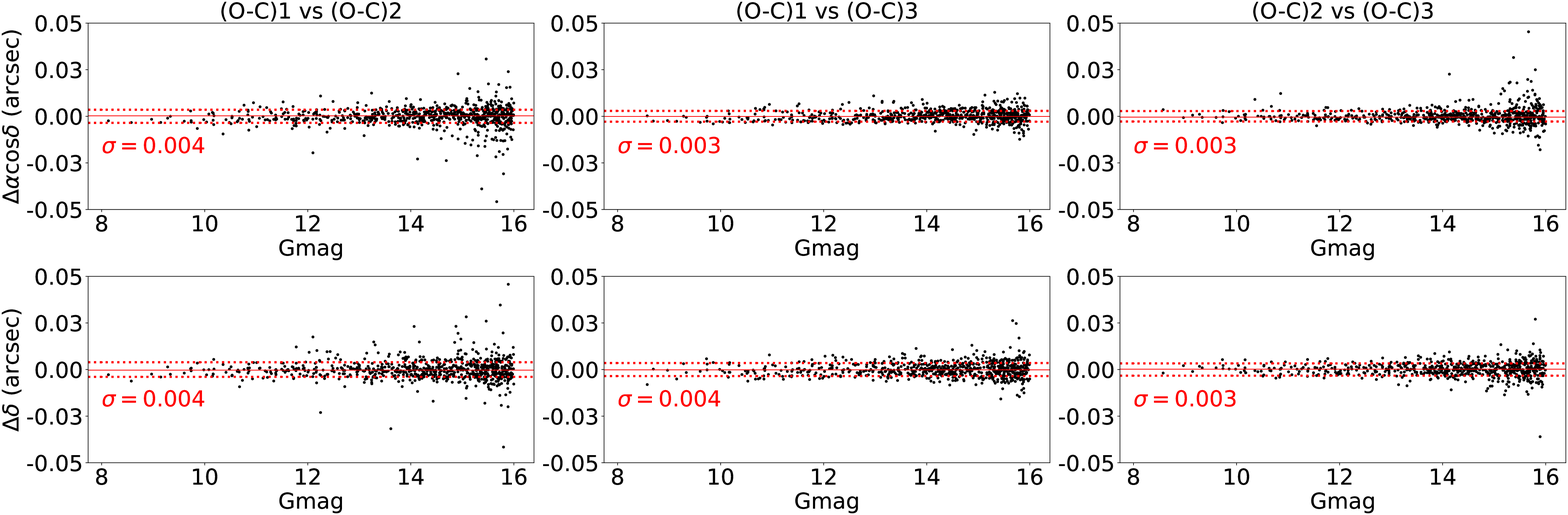} \\
\caption{The (O-C) values change between two observation sets. Left:~(O-C) values of Obs1 compare with (O-C) values of Obs2. Middle:~(O-C) values of Obs1 compare with (O-C) values of Obs3. Right:~(O-C) values of Obs2 compare with (O-C) values of Obs3. The red dashed line is marked at $\pm\sigma$ of the residuals and the solid line is marked at average value.}
\label{Fig8}
\end{figure*}

For the GD1 solution, we compare the average (O-C) of the same stars when a 6-parameter plate model is used for transformation~(see the section above), ignoring the proper motion effects of stars in one or two days. The variations of (O-C) are shown in Fig.~\ref{Fig8}. It seems that the measurement from photographic astrometry has a small advantage~($\sim$4 mas at average in one direction). The differential astrometry's measurement might be more vulnerable to environmental variations, and a higher-order polynomial is needed when cross-matching the master frame from various epochs. Based on the derived GD1 solution, we generate synthetic distorted data for further investigation in the next section.

\section{Analysis using synthetic distorted data}
In this section, we generate synthetic distorted data based on Obs3 and derive GD solutions. The use of simulation allows for the two GD solutions to be investigated free from the influence of the instrument or weather environment. And the synthetic data are generated in an imitation of the process of photographic astrometry since it is generally accepted. And what is more, the high accuracy and precision of Gaia DR2 provide a robust calculation for them.

\subsection{The generation of synthetic data and analysis of results}
Initially, we take the standard frame calculated from Gaia for each observation as ideal positions. Then each standard frame is aligned with the GD calibrated frame by conformal transformation and added with systematic uncertainties. The systematic uncertainties are based on the source's Gmag~($m$) and presented as the following sigmoidal formula in arcsecond:
$$
\begin{array}{c}
\sigma(m)=(A_1-A_2)/(1+e^{(m-m_0)/dm})+A_2
\end{array}
$$
where $A_1$ and $A_2$  are the initial and final values of the sigmoidal curve respectively, $m_0$ is the $m$ value of the curve's midpoint and $dm$ is the growth rate of the curve. We adopt the values of $A_1$~(0.0059), $A_2$~(0.183), $m_0$~(18.6) and $dm$~(0.911) which have been derived by Lin~\citep{Lin2019}. For Obs3, $A_1$ $\sim$ 0.025 pixel~(about 6 mas), serves as the systematic error floor to be expected from the synthetic data, i.e. for the positional measurement of high SNR star images.

We choose two GD1 models to be introduced into the simulation data, which are derived from the observations take at YNAO 1-m telescope~\citep{Peng2012} and 2.4-m telescope~\citep{Zheng2017} respectively. The models are expressed with quadratic and quartic polynomials respectively~($N_d$ is 2 or 4):

$$
\left\{
\begin{array}{rcl}
 \Delta x&\!\!\!=\!\!\!&  \sum\limits_{i,j,0\leqslant i+j \leqslant N_d} a_{ij}\tilde{x}^{i}\tilde{y}^{j}\\
 \Delta y&\!\!\!=\!\!\!&  \sum\limits_{i,j,0\leqslant i+j \leqslant N_d} b_{ij}\tilde{x}^{i}\tilde{y}^{j}\\
\end{array}
\right.
$$
where the input $x$ and $y$ have been normalized within $[\text{-}1,1]$ across the FOV:
$$
\left\{\begin{array}{c}
\tilde{x}=(x-2048)/2048 \\
\tilde{y}=(y-2056)/2056.
\end{array}\right.
$$The normalized positions make it easier to recognize the magnitude
of the contribution of each term at the edge of FOV.

For the GD effects introduced into the synthetic data, we only consider the non-linear distortion terms instead of the linear ones. That is, the 0- and 1-order coefficients of the introduced GD models are set as zero. The linear distortion can be absorbed by a six-parameter transformation which is easy to meet the requirement for three reference sources in practice while the gravitational flexure effect is apt to introduce a time-variable non-linear instrumental distortion~\citep{Trippe2010}. Therefore, to fully exploit the astrometry potentials, the non-linear distortion terms of several large telescopes are paid close attention to~\citep{Patti2019,Plewa2015MNRAS,Trippe2010}. Also for FITS Image's WCS~(World Coordinate System) transformation, SIP~(Simple Imaging Polynomial,~\citealt{Shupe2005}), which is used in SAOImage DS9~\citep{Joye2003}, Astrometry.net~\citep{Lang2010}, Drizzle~\citep{Fruchter2002}, adopts only the quadratic and higher-order terms of a polynomial to represent the distortion instead of the linear terms. As a distortion-free system can have arbitrary orientation and scaling, we have the freedom to choose the linear terms of our system. And the GD solution, which calibrates the observations and enables them to be transformed into the distortion-free system using only a conformal transformation, is also allowed to have arbitrary orientation and scaling. Thus, the linear terms would not be unique. \citet{Anderson2003} have used constraints as $a_{10}=0$ and $a_{01}=0$ when they solve the distortion, which means the solution has its $X$ scale equal to that of the chip and its $Y$ axis is aligned with that of the chip. We would not plan to make constraints for the linear terms, after all, they would deviate from the original GD model. We show the non-linear coefficients of the introduced GD models and the derived ones by virtue of two solutions in Table~\ref{Tab SYNGD1m}. For simplicity, the synthetic datasets introduced by two GD models are called Syn1 and Syn2 respectively.

\begin{table*}
\centering
\caption{Non-linear coefficients of the two introduced GD models and the corresponding derived GD effects by two GD solutions. The subscript ${\bot}$ denote GD solution derived from data including two perpendicular orientations. The values in red indicate that they are different from the real value up to the threshold~($0\farcs005$, $\sim$0.02 pixel).}
\begin{tabular}{c c l l l l l l l l}
\hline\hline
\multirow{2}*{Dataset} & \multirow{2}*{$ij$} & \multicolumn{2}{c}{Synthetic} & \multicolumn{2}{c}{GD1}  & \multicolumn{2}{c}{GD2} & \multicolumn{2}{c}{GD2$_{\bot}$}\\
& & \multicolumn{1}{c}{$a_{ij}$} & \multicolumn{1}{c}{$b_{ij}$} &
\multicolumn{1}{c}{$a_{ij}$} & \multicolumn{1}{c}{$b_{ij}$} &
\multicolumn{1}{c}{$a_{ij}$} & \multicolumn{1}{c}{$b_{ij}$} &
\multicolumn{1}{c}{$a_{ij}$} & \multicolumn{1}{c}{$b_{ij}$}\\
\hline
\multirow{3}*{Syn1} &\multicolumn{1}{c}{$20$}  & \multicolumn{1}{r}{-0.034} & \multicolumn{1}{r}{-0.018} & \multicolumn{1}{r}{-0.034$\pm$0.002} & \multicolumn{1}{r}{-0.019$\pm0.002$} & \multicolumn{1}{r}{\textcolor[rgb]{1.00,0.00,0.00}{-0.008}$\pm0.002$} & \multicolumn{1}{r}{\textcolor[rgb]{1.00,0.00,0.00}{0.008}$\pm0.002$} & \multicolumn{1}{r}{-0.042$\pm0.001$} & \multicolumn{1}{r}{-0.011$\pm0.001$}\\
&\multicolumn{1}{c}{$11$}  & \multicolumn{1}{r}{0.023} & \multicolumn{1}{r}{-0.091} & \multicolumn{1}{r}{0.024$\pm0.002$} & \multicolumn{1}{r}{-0.099$\pm0.002$} & \multicolumn{1}{r}{\textcolor[rgb]{1.00,0.00,0.00}{-0.011}$\pm0.002$} & \multicolumn{1}{r}{\textcolor[rgb]{1.00,0.00,0.00}{-0.049}$\pm0.002$} & \multicolumn{1}{r}{0.024$\pm0.001$} & \multicolumn{1}{r}{-0.094$\pm0.001$}\\
&\multicolumn{1}{c}{$02$}  & \multicolumn{1}{r}{-0.057} & \multicolumn{1}{r}{0.005} & \multicolumn{1}{r}{-0.051$\pm0.002$} & \multicolumn{1}{r}{0.005$\pm0.002$} & \multicolumn{1}{r}{-0.065$\pm0.002$} & \multicolumn{1}{r}{-0.004$\pm0.002$} & \multicolumn{1}{r}{-0.059$\pm0.001$} & \multicolumn{1}{r}{0.010$\pm0.001$}\\
\hline
\multirow{12}*{Syn2} & \multicolumn{1}{c}{20}  & \multicolumn{1}{r}{0.382} & \multicolumn{1}{r}{0.029} & \multicolumn{1}{r}{0.382$\pm0.007$} & \multicolumn{1}{r}{0.013$\pm0.006$} & \multicolumn{1}{r}{\textcolor[rgb]{1.00,0.00,0.00}{0.227}$\pm0.007$} & \multicolumn{1}{r}{\textcolor[rgb]{1.00,0.00,0.00}{-0.187}$\pm0.006$}
& \multicolumn{1}{r}{0.373$\pm0.005$} & \multicolumn{1}{r}{0.035$\pm0.005$} \\
&\multicolumn{1}{c}{$11$}  & \multicolumn{1}{r}{-0.312} & \multicolumn{1}{r}{0.264} & \multicolumn{1}{r}{-0.303$\pm0.006$} & \multicolumn{1}{r}{\textcolor[rgb]{1.00,0.00,0.00}{0.237}$\pm0.005$} & \multicolumn{1}{r}{\textcolor[rgb]{1.00,0.00,0.00}{0.108}$\pm0.005$} & \multicolumn{1}{r}{\textcolor[rgb]{1.00,0.00,0.00}{-0.015}$\pm0.005$}
& \multicolumn{1}{r}{-0.309$\pm0.004$} & \multicolumn{1}{r}{0.275$\pm0.004$} \\
&\multicolumn{1}{c}{$02$}  & \multicolumn{1}{r}{0.132} & \multicolumn{1}{r}{-0.294} & \multicolumn{1}{r}{0.141$\pm0.007$} & \multicolumn{1}{r}{-0.304$\pm0.006$} & \multicolumn{1}{r}{\textcolor[rgb]{1.00,0.00,0.00}{0.265}$\pm0.007$} & \multicolumn{1}{r}{\textcolor[rgb]{1.00,0.00,0.00}{-0.067}$\pm0.007$}
& \multicolumn{1}{r}{0.116$\pm0.005$} & \multicolumn{1}{r}{-0.283$\pm0.005$} \\
&\multicolumn{1}{c}{$30$} & \multicolumn{1}{r}{-0.784} & \multicolumn{1}{r}{0.013} & \multicolumn{1}{r}{-0.778$\pm0.004$} & \multicolumn{1}{r}{0.018$\pm0.003$} & \multicolumn{1}{r}{-0.773$\pm0.004$} & \multicolumn{1}{r}{0.016$\pm0.003$}
& \multicolumn{1}{r}{-0.791$\pm0.002$} & \multicolumn{1}{r}{0.015$\pm0.002$}  \\
&\multicolumn{1}{c}{$21$}  & \multicolumn{1}{r}{-0.017} & \multicolumn{1}{r}{-0.724} & \multicolumn{1}{r}{-0.023$\pm0.003$} & \multicolumn{1}{r}{-0.727$\pm0.003$} & \multicolumn{1}{r}{-0.016$\pm0.003$} & \multicolumn{1}{r}{-0.725$\pm0.003$}  & \multicolumn{1}{r}{-0.021$\pm0.002$} & \multicolumn{1}{r}{-0.726$\pm0.002$} \\
&\multicolumn{1}{c}{$12$} & \multicolumn{1}{r}{-0.724} & \multicolumn{1}{r}{-0.009}& \multicolumn{1}{r}{-0.738$\pm0.003$} & \multicolumn{1}{r}{-0.010$\pm0.003$} & \multicolumn{1}{r}{-0.730$\pm0.003$} & \multicolumn{1}{r}{-0.013$\pm0.003$} & \multicolumn{1}{r}{-0.723$\pm0.002$} & \multicolumn{1}{r}{-0.006$\pm0.002$} \\
&\multicolumn{1}{c}{$03$} & \multicolumn{1}{r}{-0.002} & \multicolumn{1}{r}{-0.736} & \multicolumn{1}{r}{0.001$\pm0.004$} & \multicolumn{1}{r}{-0.732$\pm0.003$} & \multicolumn{1}{r}{0.001$\pm0.004$} & \multicolumn{1}{r}{-0.729$\pm0.003$} & \multicolumn{1}{r}{-0.004$\pm0.002$} & \multicolumn{1}{r}{-0.741$\pm0.003$} \\
&\multicolumn{1}{c}{$40$} & \multicolumn{1}{r}{-0.015} & \multicolumn{1}{r}{-0.012} & \multicolumn{1}{r}{-0.012$\pm0.008$} & \multicolumn{1}{r}{-0.002$\pm0.007$} & \multicolumn{1}{r}{-0.011$\pm0.007$} & \multicolumn{1}{r}{-0.006$\pm0.007$}
& \multicolumn{1}{r}{-0.016$\pm0.005$} & \multicolumn{1}{r}{-0.008$\pm0.005$}  \\
&\multicolumn{1}{c}{$31$}  & \multicolumn{1}{r}{-0.100} & \multicolumn{1}{r}{0.001} & \multicolumn{1}{r}{-0.100$\pm0.006$} & \multicolumn{1}{r}{0.014$\pm0.006$} & \multicolumn{1}{r}{-0.106$\pm0.006$} & \multicolumn{1}{r}{-0.001$\pm0.006$}  & \multicolumn{1}{r}{-0.103$\pm0.004$} & \multicolumn{1}{r}{-0.008$\pm0.004$} \\
&\multicolumn{1}{c}{$22$} & \multicolumn{1}{r}{0.035} & \multicolumn{1}{r}{0.141}& \multicolumn{1}{r}{0.035$\pm0.006$} & \multicolumn{1}{r}{0.160$\pm0.006$} & \multicolumn{1}{r}{0.032$\pm0.006$} & \multicolumn{1}{r}{0.154$\pm0.006$} & \multicolumn{1}{r}{0.037$\pm0.004$} & \multicolumn{1}{r}{0.136$\pm0.004$} \\
&\multicolumn{1}{c}{$13$} & \multicolumn{1}{r}{-0.118} & \multicolumn{1}{r}{0.005} & \multicolumn{1}{r}{-0.126$\pm0.007$} & \multicolumn{1}{r}{0.017$\pm0.006$} & \multicolumn{1}{r}{-0.119$\pm0.006$} & \multicolumn{1}{r}{0.014$\pm0.006$} & \multicolumn{1}{r}{-0.119$\pm0.004$} & \multicolumn{1}{r}{-0.009$\pm0.004$} \\
&\multicolumn{1}{c}{$04$} & \multicolumn{1}{r}{-0.013} & \multicolumn{1}{r}{-0.077} & \multicolumn{1}{r}{-0.016$\pm0.008$} & \multicolumn{1}{r}{-0.074$\pm0.007$} & \multicolumn{1}{r}{-0.012$\pm0.007$} & \multicolumn{1}{r}{-0.076$\pm0.007$} & \multicolumn{1}{r}{0.002$\pm0.005$} & \multicolumn{1}{r}{-0.082$\pm0.005$} \\
\hline
\end{tabular}
\label{Tab SYNGD1m}
\end{table*}

We set a threshold of $0\farcs005$~($\sim$0.02 pixel) to evaluate the derived coefficients. The results derived by GD1 solution seem to have a nice agreement with the synthetic GD effect, although for Syn2, $b_{11}$ has the largest coefficient difference, 0.027 pixel. For GD2, quite big discrepancies are found in most quadratic coefficients especially for Syn2 data, but the cubic and quartic coefficients approximate the real values.

As indicated by \citet{Anderson2003}, the linear terms of GD2 solutions can be derived correctly from observations at various orientations. Inspired by this idea, additional synthetic data are generated, which are perpendicular to the original observations, while the original synthetic dataset is kept without losing the accuracy and precision of the master frame. Note that each exposure of the newly added data is rotated anticlockwise by 90 degree with respect to the original observation. It is found that all of the quadratic coefficients are close to the real values within the threshold~(the last two columns in Table~\ref{Tab SYNGD1m}). Therefore we come to the conclusion that the quadratic terms of GD2 solution are degenerated with the solution for inter-image offsets, if the observations are at the same orientation.

We think it is similar to the situation of the linear terms, which are easily coupled with DFF's positions. As mentioned above, DFF is allowed to have arbitrary orientation and scaling. Therefore it is not necessarily the same as the real location. The supposed ideal positions must affect the GD solution, especially for the low-order coefficients. For a ground-based telescope that has many high SNR sources to work with, most of which are fainter than Gaia's limit, GD2 solution might be a better option. But more optimally dithered, overlapping frames at different orientations may be required to ensure the accuracy of the linear and quadratic terms of GD2 solution. And many ground-based telescopes with rigid mounting of the telescope itself and of its detectors, are unable to achieve observations at various orientations~\citep{Anderson2003}. In this case, at least six reference stars are needed to construct a second-order polynomial after GD2 solution, in order to absorb the residual distortion.

\subsection{The relation between the standard frame and the master frame}
In this section, we discuss the relationship between the standard frame and the master frame. The best representative of the master frame should be the HST astrometric reference catalogs, due to their extremely high precisions. Among them, yearly observations of 47 Tuc have been taken over the lifetime of the HST to characterize and correct its distortion to provide high precision astrometric reference catalogs~\citep{Borncamp2015AAS}. However, the astrometric reference catalog is found to have measurable differences in rotation and scale with respect to Gaia DR1~\citep{2018wfc}.

Since the observations' $X$ and $Y$ directions are not necessarily aligned with R.A. and Decl. directions respectively, the rotation between the master frame and ICRS is allowed. But the variations of the scale across the FOV will bias relative positioning such as the determination for relative proper motions of cluster members. The cause for the variation of the scale is that, observations for construction of the master frame have different tangential points. If they are bonded to each other directly in the pixel coordinate without considering a gnomonic projection, the plate scale of the master frame will vary across the FOV.

Nonetheless, in the time before the release of Gaia, HST astrometric reference catalogs still serve as useful references for many cutting edge examples of research in astronomy. For example, the HST observations of M92 are taken as a distortion-free reference frame to improve the GD solution of Keck II 10-m telescope's near-infrared camera~(NIRC2) in its narrow field mode, which is a major limitation for the proper motion measurements of Galactic central stellar cluster~\citep{Yelda2010}. And recently, based on Gaia DR2's positions, not only the GD solution of HST but the astrometric reference catalog of 47 Tuc are improved in terms of precision and accuracy~\citep{Hoffmann2020}.

\section{Conclusion} \label{sec:conclusion}

Following the second release of Gaia catalog, photographic astrometry will be expected to become more widespread. Before that, most of the large telescopes perform differential astrometry only using pixel/measurement coordinates of the image since few sources in the FOV would have the celestial information from the catalog. Both types of astrometry need careful calibration especially for GD effects in the FOV. For two types of astrometry, GD solution can be derived based on either the positions calculated from an astrometric catalog or the positions of the master frame, which is constructed from observations themselves.

Taking advantage of Gaia DR2, we perform the two GD solutions for the observations taken with 1-m telescope at YNAO to study their discrepancies. For the same observations, the GD patterns derived from the two solutions are different. Further analysis is done for the two solutions in terms of measurement precision and accuracy. The (internal) precisions after the two GD corrections agree well with each other. However, when we compare positions of different dates, derived from photographic astrometry or differential astrometry, it is found that photographic astrometry provides more reliable positions.

To avoid the influence of the instrument and weather environment, a simulation is performed from the sight of photographic astrometry for further investigation. After two GD models are derived from practical observations, the effects of their non-linear terms are added to the synthetic pixel positions of the standard frame calculated from Gaia DR2 and then the two GD solutions are performed. The results of the GD1 solution have a nice agreement with the introduced GD models. For the GD2 solution, the quadratic terms can not be solved for if the observations are at the same orientation while higher-order terms approach the real values. This indicates that a second-order polynomial is needed to absorb the residual distortion when the observations are in the same orientation. In practice, more optimally dithered, overlapping frames at different orientations may be required to ensure the success of self-calibration. However, many ground-based telescopes with rigid mounting of the telescope itself and of its detectors, are unable to achieve them~\citep{Anderson2003}.

For wide-field projects such as surveys or proper motion solution, it might be a better option to implement astrometric GD calibration since the Gaia catalog offers high-accuracy positions and proper motions in this decade. Although still awaiting the final Gaia catalog, now Gaia DR2 suffices to act as a concrete reference frame for the reliable astrometric calibration for new generation telescopes such as the Dark Energy Camera~\citep{Bernstein2017}. And as in the golden age of space astrometry, in the near future space missions will become active and ambitious, driving the successive advent of high-accuracy catalogs, and photographic astrometric reduction will play a role as a touchstone for the project quality for large imaging telescopes.

\section*{Acknowledgements}

This work was supported by the National Natural
Science Foundation of China (Grant Nos.~11873026,~11703008,~11273014), by
the Joint Research Fund in Astronomy (Grant No. U1431227) under
cooperative agreement between the National Natural Science
Foundation of China~(NSFC) and Chinese Academy Sciences~(CAS), and
partly by the Fundamental Research Funds for the Central
Universities. We thank the referee for the insightful review of our manuscript. We also thank Dr. Nick Cooper in Queen Mary University of London for his linguistic assistance during the preparation of this manuscript. We also thank the chief scientist Qian S.B. of the 1-m telescope and his working group for their
kindly support and help. This work has made use of data from the European Space Agency (ESA) mission \emph{Gaia} (\url{https://www.cosmos.esa.int/gaia}), processed by the \emph{Gaia} Data Processing and Analysis Consortium (DPAC, \url{https://www.cosmos.esa.int/web/gaia/dpac/consortium}). Funding for the DPAC has been provided by national institutions, in particular the institutions participating in the \emph{Gaia} Multilateral Agreement.

\section*{DATA AVAILABILITY}
The data underlying this article will be shared on reasonable request
to the corresponding author.



\bibliographystyle{mnras}
\bibliography{mypaper} 




%
%


\bsp	
\label{lastpage}
\end{document}